\documentclass[epj]{svjour}
\usepackage[english]{babel}
\usepackage{graphicx}
 \usepackage{amsmath}
 \usepackage{amsfonts}
 \usepackage{amssymb}
 \usepackage{pstricks}
 \usepackage{psfrag}
 \usepackage{epsfig}
 \usepackage{color}
 \usepackage[ansinew]{inputenc}
 \usepackage{cite}

\begin{document}

\title{
Systematic extension of the Cahn-Hilliard model for motility-induced phase separation
}

\author{Lisa Rapp \and Fabian Bergmann \and Walter Zimmermann}
\authorrunning{Rapp et al.}

\institute{Theoretische Physik I, Universit\"at Bayreuth, 95440 Bayreuth, Germany}

\date{Received: \today / Revised version: (date)}

\abstract{ We consider a continuum model for  motility-induced phase separation
(MIPS) of active Brownian particles [J. Chem. Phys. {\bf 142}, 224149 (2015)].
Using a recently introduced perturbative analysis [Phys. Rev. E {\bf
98}, 020604(R) (2018)], we show that this continuum model reduces 
 to the classic Cahn-Hilliard (CH) model near the onset of MIPS. 
This makes MIPS another example of the so-called active phase separation.
We further introduce a generalization of the perturbative analysis to
the next higher order.
This results in a generic higher order extension of the CH model for active phase separation. 
 Our analysis establishes the mathematical link between the basic mean-field MIPS model on the one hand,
and the leading order and extended CH models on the other hand. 
Comparing numerical simulations of the three models, we find that 
the leading order CH model agrees nearly perfectly with the full continuum model near the onset of MIPS. 
We also give estimates of the control parameter beyond which the higher order corrections become relevant and 
compare the extended CH model to recent phenomenological models.
}

\maketitle

%%%%%%%%%%%%%%%%%%%%%%%%%%%%%%%%%%%%%%%%%%%%%%%%%%%%%%%%%%%%%%%%%%%%%%%%%%%%%%%
%      SECTION:  INTRODUCTION
%%%%%%%%%%%%%%%%%%%%%%%%%%%%%%%%%%%%%%%%%%%%%%%%%%%%%%%%%%%%%%%%%%%%%%%%%%%%%%%
\section{Introduction}\label{sec: intro}

Active matter systems are nonequilibrium systems which consume fuel and disspative 
energy locally. These systems 
are full of fascinating  phenomena and have recently attracted increasing
attention in the scientific community  \cite{Vicsek:2012.1,Giardina:2014.1,HymanA:2014.1,Cates:2015.1,Aranson:2015.1,Bechinger:2016.1,Brangwynne:2017.1,Juelicher:2018.1}.
Examples range 
from active molecular processes
which are driven by chemical free energy provided by metabolic processes \cite{Alberts:2001}
up to flocks of birds and schools of fish \cite{Vicsek:2012.1,Giardina:2014.1}.
Various active matter systems also show collective non-equilibrium transitions.
On the time scale of these transitions, the involved entities such as proteins, cells or even birds 
are conserved. Examples include cell polarization  
\cite{EdelsteinKeshet:2007.1,Kuroda:2007.1,Grill:2013.1,Grill:2014.1,Alonso:2010.1,GoehringNW:2011.1,Bergmann:2018.2},
chemotactically communicating cells \cite{Tindall:2008.1,HillenT:2009.1,Romanczuk:2014.1,Liebchen:2015.1}, 
self-propelled colloidal particles \cite{Theurkauff:2012.1,Palacci:2013.1,SpeckT:2013.1,Marchetti:2012.1,Cates:2013.1,Baskaran:2013.2,SpeckT:2014.1},
as well as mussels in ecology \cite{vandeKoppel:2013.1}.

Self-propelling colloidal particles undergo a non-equilibrium phase transition
into two distinct phases - a denser liquid-like phase and a dilute gas-like phase
\cite{Theurkauff:2012.1,Palacci:2013.1,SpeckT:2013.1} -
if their swimming speed decreases with increasing local density. 
This is known as motility-induced phase separation (MIPS)  \cite{Marchetti:2012.1,Baskaran:2013.2,Cates:2015.1}.
It strikingly resembles well-known phase separation processes at thermal equilibrium such as the demixing of a binary fluid.
We recently introduced a class of such non-equilibrium demixing phenomena  we call active phase separation \cite{Bergmann:2018.2}.
Among the phenomena identified as members of this class are cell polarization or chemotactically communicating cells.
For this class we have shown that the similarities between equilibrium and nonequilibrium demixing phenomena are 
in fact not coincidental. We have generalized 
a classical weakly nonlinear analysis  near a supercritical bifurcation with unconserved order parameter fields
\cite{CrossHo} to the case of active phase separation with a conserved order parameter field \cite{Bergmann:2018.2}.
The generic equation describing active phase separation systems turned out to be the classic Cahn-Hilliard  (CH) model 
- the same generic model that also describes equilibrium phase separation.

In this work, we raise the question whether the recently introduced nonlinear perturbation approach in Ref.~\cite{Bergmann:2018.2}
is also directly applicable to MIPS. 
We employ this reduction approach to a mean-field description of MIPS provided by Speck et al \cite{SpeckT:2014.1,SpeckT:2015.1} 
and show how the MIPS model reduces to the CH model at leading order.  

Recently, several phenomenological extensions of the CH model have also been 
considered as continuum models of MIPS \cite{Cates:2014.1,Cates:2018.1}. 
These are extensions of the CH model to the next higher order of
nonlinear contributions.  
In this work, we therefore also introduce an extension of our perturbative scheme 
that allows us to systematically derive higher order nonlinearities directly from the continuum model for MIPS.
Due to our systematic approach, the extended CH model we derive is not 
 a phenomenological model. 
Instead, we directly map the continuum model for MIPS to the extended CH model. 
Note that we concentrate on the example of MIPS in this work.
However, the extension introduced here can be applied to any system in the class of active phase separation. 
We thus show in general how both the leading order CH model and its 
extension describe active phase separation as a non-equilibrium phenomenon.

This work is organized as follows:
We first present the mean-field MIPS model and calculate the onset of phase
separation in the system.
We then introduce the perturbative scheme we use to reduce
the MIPS model to the classic CH equation near the onset of phase separation. 
In the next step, we extend the previous approach to include
 nonlinearities at the next higher order.
Section 5 is an in-depth discussion of the derived leading order and extended CH models including their connection to the mean-field MIPS model and other
phenomenological descriptions of MIPS. 
Finally, in section 6, we present numerical simulations comparing leading order and extended CH to the full mean-field MIPS model 
to assess validity and accuracy of the reduced models.

%%%%%%%%%%%%%%%%%%%%%%%%%%%%%%%%%%%%%%%%%%%%%%%%%%%%%%%%%%%%%%%%%%%%%%%%%%%%%%%
%      SECTION:  MODEL
%%%%%%%%%%%%%%%%%%%%%%%%%%%%%%%%%%%%%%%%%%%%%%%%%%%%%%%%%%%%%%%%%%%%%%%%%%%%%%%

\section{Model}\label{sec: model} 

On a mean-field level, phase separation of active Brownian particles
can be described by two coupled equations for the 
particle density $\tilde \rho({\bf r},t)$ and a polarization ${\bf p}({\bf r},t)$ \cite{SpeckT:2013.1,SpeckT:2015.1}.
The evolution of the  particle density $\tilde \rho$ is determined by
\begin{equation}
 \partial_t \tilde \rho = -\nabla\cdot\left[v(\tilde \rho)\mathbf{p}-D_e\nabla \tilde \rho\right]\,,
 \label{eq:density}
\end{equation}
where $D_e$ is the effective diffusion coefficient of the active Brownian particles.
 $v(\tilde \rho)$ is the density-dependent particle speed given by
\begin{equation}
 v(\tilde \rho) = v_0 -\tilde \rho \zeta +\lambda^2\nabla^2 \tilde \rho \,.
 \label{eq:effective_speed}
\end{equation}
$v_0$ is the speed of a single self-propelled particle. 
With  increasing particle density, the velocity is reduced by $\zeta \tilde \rho$ due to interactions with other particles.
$\zeta$ is related to the pair distribution function of the individual particles and
 assumed to be spatially homogeneous \cite{SpeckT:2013.1}. The  nonlocal contribution in eq.~({\ref{eq:effective_speed})
was introduced in Refs.~\cite{Cates:2013.1,Cates:2014.1}. 
It incorporates the effect that active Brownian 
particles sample the neighboring particle density on a length scale $\lambda$ larger than the particle spacing.
Equation (\ref{eq:effective_speed}) 
is coupled to a dynamical equation for the polarization \cite{SpeckT:2013.1,SpeckT:2015.1},
\begin{equation}
 \partial_t\mathbf{p} = -\nabla P(\tilde \rho) + D_e\nabla^2\mathbf{p}-\mathbf{p},
 \label{eq:polarization}
\end{equation}
with the ``pressure'' 
\begin{equation}
 P(\tilde \rho) = \frac{1}{2}v(\tilde \rho)\tilde \rho\,.
\end{equation}

\section{Onset of phase separation}
\label{sec:linsta}

The stationary solution of eq.~(\ref{eq:density}) and eq.~(\ref{eq:polarization}) is any constant density $\bar \rho$
and $\mathbf{p}=0$. Therefore, 
 we decompose the particle density into its homogeneous part $\bar \rho$ and the inhomogeneous density variation $\rho$:
\begin{equation}
 \tilde \rho = \bar{\rho} + \rho \,.
\end{equation}
Accordingly, we investigate the following dynamical equations for $ \rho$ and ${\bf p}$:
\begin{subequations}
\label{sepbasic}
\begin{align}
 \partial_t \rho &= -\nabla \left[ \alpha -\zeta  \rho +\lambda^2\nabla^2  \rho \right] {\bf p}  +D_e \Delta  \rho\,, \\
 \partial_t {\bf p} & = -\nabla \left[ \beta  \rho -\frac{1}{2} \zeta {\rho}^2 +\frac{\lambda^2}{2} \left(\bar \rho +
 \rho \right)
 \nabla^2  \rho \right] \nonumber \\
                    & \qquad +D_e \Delta {\bf p} -{\bf p}\,,
\end{align}
\end{subequations}
where
\begin{align}
 \alpha =v_0-R, \qquad \beta=\frac{1}{2}(v_0-2 R)\,,
\end{align}
with the density parameter
\begin{align}
 R=\zeta \bar  \rho\,.
\end{align}
We assume $\zeta$ and $D_e$ to be constant \cite{SpeckT:2015.1}.

The homogeneous basic solution $ \rho=0,~ {\bf p}=0$ 
is unstable if the perturbations 
$ \rho, {\bf p} = \hat \rho, \hat {\bf p} \exp(\sigma t + iqx)$ grow, 
{\it i.e.} if the growth rate $\sigma$ is positive.
Solving the linear parts of eqs.~(\ref{sepbasic}) with this perturbation ansatz, we find the dispersion relation 
\begin{align}
\label{basicdisp}
 \sigma(q)&=-\frac{1}{2} -D_eq^2 +\frac{1}{2}\sqrt{1-4 \alpha \beta q^2+2\lambda^2\alpha\rho_0q^4}\,, \nonumber \\
          &= D_2q^2 - D_4q^4 + {\cal O}(q^6)\,,
\end{align}
where 
\begin{align}
\label{eq:full_D2}
 D_2 &= -(D_e+\alpha \beta)\,,\\
 D_4 &= \left(\alpha^2 \beta^2-\frac{\lambda^2}{2}\frac{R}{\zeta}\alpha\right). 
 \label{eq:full_D4}
\end{align}
$D_2$ changes its sign as a function of $v_0$.
Assuming $D_4>0$, the growth rate $\sigma$ becomes positive in a finite range of $q=[0,q_{max}]$,
when $D_2>0$. Note that the range of wavenumbers $q$ with positive growth rate extends down to $q=0$.
The related instability condition
\begin{equation}
 D_e+\alpha\beta=0
 \label{eq:instability_condition}
\end{equation}
provides a quadratic polynomial for the critical mean density $\bar\rho$ 
(represented by the density parameter $R$) and the respective particle speed $v_0(R)$:
\begin{equation}
 \frac{1}{2}v_0^2-\frac{3}{2}Rv_0 +D_e+R^2=0.
\end{equation}
For particle speeds $v_0>v_\ast$, where
\begin{align}
 v_* &= 4\sqrt{D_e}\,,
 \label{eq:vstar_def}
\end{align}
this polynomial  has two real  solutions
\begin{equation}
 R_{\pm}=\frac{1}{4}\left[3v_0 \pm \sqrt{v_0^2-16D_e}\right]\,.
 \label{eq:Rpm_def}
\end{equation}
This corresponds to a critical value $R_*$ of the density parameter:
\begin{align}
 R_*&=R(v_*)=\frac{3}{4} v_*\,.
 \label{eq:Rstar_def}
\end{align}
Note that the assumption $D_4>0$ is fulfilled if $\lambda^2<2\zeta\alpha\beta^2/R$,
{\it i.e.} for sufficiently small $\lambda$.
At the critical point, $v_0=v_*$ and $R=R_*$, this condition simplifies to
\begin{equation}
\lambda^2<\zeta v_*^2/24.
\label{eq:lambda_condition}
\end{equation}
For particle velocities below  $v_\ast$, the homogeneous solution is stable for any value of the density parameter $R=\zeta \bar \rho$. 
For $v>v_\ast$ and $R_-<R<R_+$  (shaded region in fig.~\ref{fig1}) the homogeneous 
particle density  becomes unstable with respect to perturbations.
\begin{figure}
 \includegraphics[width=\columnwidth]{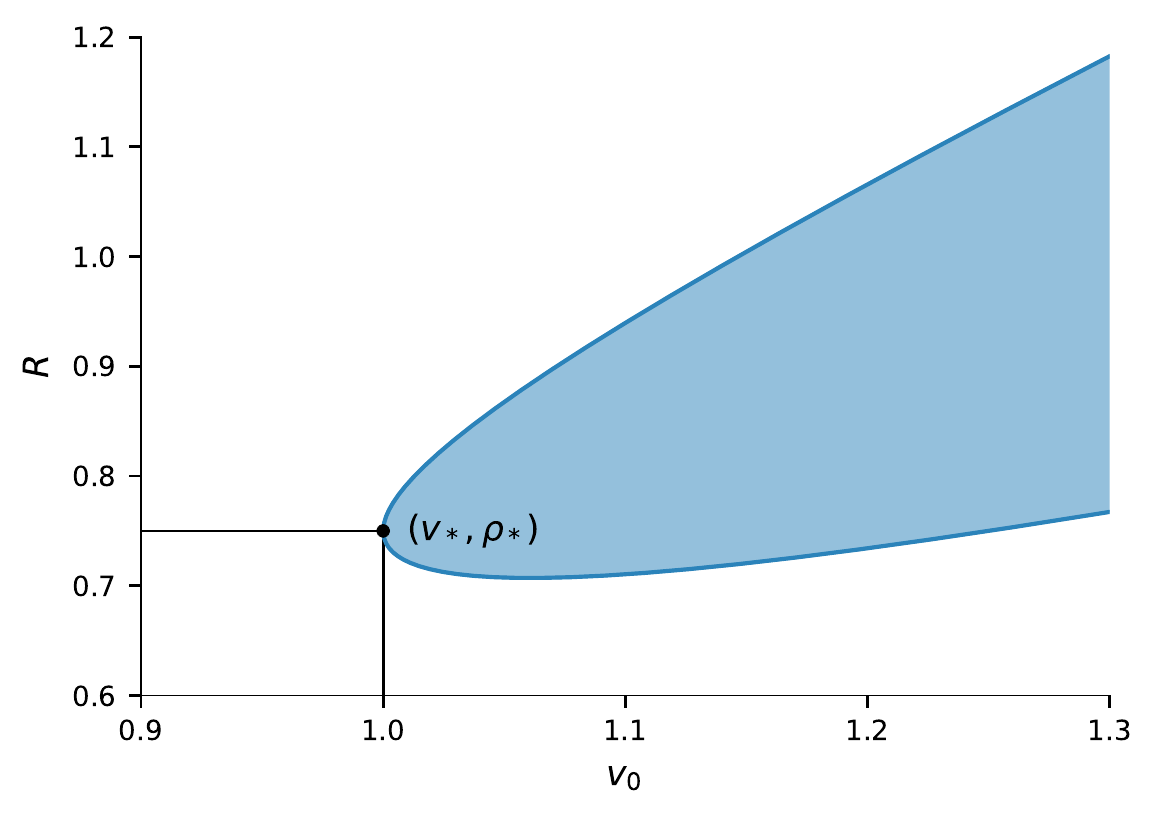}
 \caption{Instability curve $R_{\pm}(v_0)$ as given by eq.~(\ref{eq:Rpm_def}). 
 The minimum of the parabolic function is at $(v_*,R_*)=(1.0,0.75)$,
 assuming $\zeta=1$, $D_e=1/16$.
 For $v_0>v_*$, the homogeneous solution is unstable for mean densities within the shaded region.
 }
 \label{fig1}
\end{figure}

%%%%%%%%%%%%%%%%%%%%%%%%%%%%%%%%%%%%%%%%%%%%%%%%%%%%%%%%%%%%%%%%
\section{Derivation of Cahn-Hilliard models 
}
%%%%%%%%%%%%%%%%%%%%%%%%%%%%%%%%%%%%%%%%%%%%%%%%%%%%%%%%%%%%%%
\label{sec:herleitung}

In this chapter, we will apply the systematic pertubative scheme introduced recently in Ref.~\cite{Bergmann:2018.2}
to the mean field model, eqs.~(\ref{sepbasic}),
and reduce them near onset to the well-known Cahn-Hilliard (CH) model.
In a second step, we will then expand the pertubative scheme 
to include higher order contributions.

The transition from the homogenous state of eqs.~\eqref{eq:density} and \eqref{eq:polarization} 
to MIPS is either supercritical or slightly subcritical.
In both cases, cubic nonlinearities limit the growth of density modulations - as we also confirm in this work a posteriori.
Therefore, the amplitudes of the density modulations near MIPS are small and we write
\begin{align}
\label{basicscale}
 \rho=\sqrt{\varepsilon} \rho_1
\end{align}
with a small parameter $\varepsilon$ and $\rho_1 \sim {\it O}(1)$. 
Thereby $\varepsilon$ measures the distance from the critical velocity $v_\ast$:
\begin{align}
\label{v0exp}
 v_0=v_\ast(1+\varepsilon)\,.
\end{align}
This also allows an expansion of $R_\pm(v_0)$ in eq.~(\ref{eq:Rpm_def}) near $R_\ast$. At leading order, we find
$ R_\pm\simeq R_*(1 \pm \gamma_1 \sqrt{\varepsilon})$
with $ \gamma_1=\sqrt{2}/3$.
This suggests the following parameterization of $R$ in the ranges $v_0>v_\ast$ and $R_-<R<R_+$  near $R^\ast$:
\begin{align}
 R=R_\ast(1+r_1)\,, ~~~~~ \mbox{with} \qquad r_1=\sqrt{\varepsilon}\tilde r_1\,.
 \label{eq:r1_def}
\end{align}

According to the dispersion relation in eq.~(\ref{basicdisp}), 
the fastest growing mode is given by $q_e^2=D_2/(2D_4)$. 
The largest growing wavenumber $q_{max}$ (calculated from $\sigma=0$) is $q_{max}^2=D_2/D_4$. 
Thus, both $q_e^2$ and $q_{max}^2$ scale with the factor $D_2/D_4$.
Using the previously introduced definitions and expanding for small values of the control parameter $\varepsilon$,
we find $D_2/D_4\propto\varepsilon$ at leading order.
Thus, both $q_e$ and $q_{max}$ are of the order $\sqrt{\varepsilon}$, {\it i.e.} perturbations of the homogeneous basic state vary on a large length scale.
Accordingly, we introduce the new scaling $\tilde x =\sqrt{\varepsilon} x$,
resulting in the following replacement of the differential operator:
\begin{align}
 \partial_x \rightarrow \sqrt{\varepsilon}\tilde \partial_x\,.
\end{align}
From $q^2$ of order $\mathcal{O}(\varepsilon)$ and $D_2\propto \varepsilon$ follows that $\sigma \propto \varepsilon^2$ according to eq.~(\ref{basicdisp}). 
Thus, the growth of these long wavelength perturbations is very slow.
Accordingly, we introduce the slow time scale $T_1=\varepsilon^2t$. 
In order to capture the dynamics at the next higher order, we also introduce a second slow time scale $T_2=\varepsilon^{5/2}t$. 
This suggests the following replacement of the time derivatives:
\begin{equation}
 \partial_t \rightarrow \varepsilon^2\partial_{T_1} + \varepsilon^{5/2}\partial_{T_2}.
\end{equation}
Since we expressed the density $\rho$ as a multiple of $\sqrt{\varepsilon}$, see eq.~(\ref{basicscale}), 
we also expand the polarization field ${\bf p}$ in orders of $\sqrt{\varepsilon}$:
\begin{align}
 \mathbf{p} &= \sqrt{\varepsilon}\mathbf{p_0}+\varepsilon\mathbf{p_1} + \varepsilon^{3/2}\mathbf{p_2}+\varepsilon^2\mathbf{p_3}+ 
 \varepsilon^{5/2}\mathbf{p_4}+  ...
\end{align}

We insert these scalings into the dynamic equations (\ref{sepbasic})
and collect terms of the same order $\sqrt{\varepsilon}^n$.
The polarization follows the density field adiabatically. 
Thus, the contributions to the polarization in increasing orders up to $\varepsilon^{5/2}$ are:
\begin{align}
\mathbf{p_0} &= 0,\\
 \mathbf{p_1} &= -\beta_*\tilde\partial_x\rho_1,\\
 \mathbf{p_2} &= R_*\tilde r_1 \tilde\partial_x\rho_1 + \frac{\zeta}{2}\tilde\partial_x(\rho_1^2),\\
 \mathbf{p_3} &= -\frac{v_*}{2}\tilde\partial_x\rho_1 - \left(D_e\beta_* + \frac{\lambda^2}{2}\frac{R_*}{\zeta}\right)\tilde\partial_x^3\rho_1,\\
 \mathbf{p_4} &= D_e\tilde\partial_x^3\left(\tilde r_1R_*\rho_1+\frac{\zeta}{2}\rho_1^2\right)\nonumber\\
 & \quad- \frac{\lambda^2}{2}\tilde\partial_x\left(\tilde r_1\rho_*+\rho_1\right)\tilde\partial_x^2\rho_1.
\end{align}
With these solutions, we can systematically solve the equations for the density $\rho_1$ in the successive orders of $\sqrt{\varepsilon}$.
In the lowest order $\mathcal{O}(\varepsilon^{3/2})$, we find
\begin{equation}
 0 = \left(\alpha_*\beta_*+D_e\right)\tilde\partial_x^2\rho_1.
\end{equation}
This equation, however, is trivially satisfied due to the instability condition ${\alpha_*\beta_*+D_e=0}$.

At order $\mathcal{O}(\varepsilon^2)$, we get
\begin{align}
\label{GlrhoO1}
0 &= - (\alpha_*+\beta_*)\left[ R_*\tilde{r}_1 \tilde\partial_x^2 \rho_1 + \zeta\tilde\partial_x \left(\rho_1\tilde\partial_x\rho_1\right)\right]\,.
\end{align}
With the definition of $R_*$ in eq.~(\ref{eq:vstar_def}) it follows that $\alpha_*+\beta_*=0$.
Thus, eq.~(\ref{GlrhoO1}) is again trivially fulfilled.

At order $\mathcal{O}(\varepsilon^{5/2})$, we finally get a dynamic equation for $\rho_1$:
\begin{align}
 \partial_{T_1}\rho_1 = -\tilde\partial_x^2&\left[\left(\frac{1}{8}v_*^2-\frac{9}{16}v_{*}^2\tilde r_1^2\right)\rho_1\right.\nonumber\\
 & + \left.\left(\frac{1}{256}v_{*}^4-\frac{3}{32\zeta}\lambda^2v_{*}^2\right)\tilde\partial_x^2\rho_1\right.\nonumber\\
 & - \left.\frac{3}{4}\zeta v_*\tilde r_1\, \rho_1^2 - \frac{1}{3}\zeta^2\,\rho_1^3\right]\,.
 \label{eq:ampgl_firstorder}
\end{align}
Note that we used the expressions in eq.~(\ref{eq:vstar_def}) to eliminate $R_*$ and $D_e$.
Equation~(\ref{eq:ampgl_firstorder}) has the form of the well-known Cahn-Hilliard (CH) equation \cite{Bray:1994.1,Desai:2009}.
This shows that  MIPS is a further example of the non-equilibrium demixing phenomenon which shares  the universal
CH model with classic phase separation. 
Recently, the notion {\it active phase separation} was coined for these types of non-equilibrium phenomena \cite{Bergmann:2018.2}.
Other recently discussed examples of active phase separation are cell polarization 
or chemotactically communicating cell colonies \cite{Bergmann:2018.2}.
All of these very different systems can be reduced to the same universal equation near the onset of phase separation.
They thus share generic features as expressed in their common representation via the CH equation.\\

In the next step, we extend the reduction scheme introduced in Ref.~\cite{Bergmann:2018.2} to 
include higher order nonlinearities. 
Continuing the expansion above to the next order $\mathcal{O}(\varepsilon^3)$, we obtain:
\begin{align}
 \partial_{T_2}\rho_1 = -\tilde\partial_x^2&\left[ \frac{9}{8}v_*^2\tilde r_1 \rho_1 + \frac{3}{16\zeta}\lambda^2v_*^2\tilde r_1 \left(\tilde\partial_x^2\rho_1\right)+ \frac{3}{4}\zeta v_* \rho_1^2\right.\nonumber\\
 & + \left(\frac{3}{128}\zeta v_{*}^3-\frac{5}{16}\lambda^2v_{*}\right)\left(\tilde\partial_x\rho_1\right)^2 \nonumber\\
 & + \left.\frac{\lambda^2}{8}v_{*}\tilde\partial_x^2\rho_1^2\right] \,.
 \label{eq:ampgl_secondorder}
\end{align}
We will discuss these new contributions in detail in Chapter \ref{sec:discussion_higherorder} below.

Equations~(\ref{eq:ampgl_firstorder}) and (\ref{eq:ampgl_secondorder}) can be combined 
into a single equation by reconstituting the original time scale 
via $\partial_t \rho_1 = \varepsilon^{2}\partial_{T_1}\rho_1 + \varepsilon^{5/2}\partial_{T_2}\rho_1$.
In addition, we go back to the original spatial scaling by setting $\tilde\partial_x = \partial_x/\sqrt{\varepsilon}$,
to the original density $\rho$ via eq.~(\ref{basicscale}), and $r_1$ as defined in eq.~(\ref{eq:r1_def}).
The complete extended amplitude equation for the density variations $\rho$ then reads:
\begin{align}
 \partial_t\rho = -\partial_x^2&\left[\left(\alpha_1+\beta_1\right)\rho+\left(\alpha_2+\beta_2\right)\partial_x^2\rho\right.\nonumber\\
 &+\left(\alpha_3+\beta_3\right)\rho^2-\alpha_4\rho^3\nonumber\\
 &+\left. \beta_5 \left(\partial_x\rho\right)^2 + \beta_6\partial_x^2\rho^2\right]\,.
 \label{eq:ampgl_full_short}
\end{align}
In this equation, contributions with the coefficients $\alpha_i$ originate from the leading order and are given by
\begin{subequations}
\label{eq:alpha_g}
\begin{align}
 \alpha_1 &= \frac{1}{8}v_{*}^2\varepsilon-\frac{9}{16}v_{*}^2r_1^2\,,\label{eq:alpha1}\\
 \alpha_2 &= \frac{1}{256}v_{*}^4-\frac{3}{32\zeta}\lambda^2v_{*}^2\,, \\
 \alpha_3 &= -\frac{3}{4}\zeta v_{*}r_1\,,\\
 \alpha_4 &= \frac{1}{3}\zeta^2\,.\label{eq:alpha4}
\end{align}
\end{subequations}
In other words, eq.~(\ref{eq:ampgl_full_short}) with $\beta_i=0$ is the rescaled version of eq.~(\ref{eq:ampgl_firstorder}).
The coefficients $\beta_i$ signal the new contributions from the next higher order.
They are given by
\begin{subequations}
\label{eq:beta_i}
\begin{align}
 \beta_1 &= \frac{9}{8}v_{*}^2 r_1\varepsilon\,,\label{eq:beta1}\\
 \beta_2 &= \frac{3}{16\zeta}\lambda^2v_{*}^2r_1\,,\label{eq:beta2}\\
 \beta_3 &= \frac{3}{4}\zeta v_{*}\varepsilon\,,\label{eq:beta3}\\
 \beta_5 &= \frac{3}{128}\zeta v_{*}^3-\frac{5}{16}\lambda^2v_{*}\,,\label{eq:beta5}\\
 \beta_6 &= \frac{\lambda^2}{8}v_{*}\,.\label{eq:beta6}
\end{align}
\end{subequations}

\section{Discussion of the derived Cahn-Hilliard models}

In this section, we will discuss the results obtained in the previous section \ref{sec:herleitung}. 
At first we consider the classic CH equation that resulted at leading order of our perturbative analysis.
We then  take a closer look at the  higher order corrections $\propto \beta_i$ in eq.~(\ref{eq:ampgl_full_short}). 
We also focus 
on the relation of the higher order coefficients $\beta_i$
to the parameters of recently introduced phenomenological extensions of the CH model for MIPS \cite{Cates:2014.1,Solon:2018.1,Cates:2018.1}.
%including a discussion of some special cases, and how the coefficients 

%%% Discussion of lowest order terms
\subsection{Classic CH equation at leading order}
\label{sec:discussion_leading}

For $\beta_i=0$, the leading order of eq.~(\ref{eq:ampgl_full_short}),
\begin{equation}
 \partial_t\rho = -\partial_x^2\left[\alpha_1\rho+\alpha_2\partial_x^2\rho+\alpha_3\rho^2-\alpha_4\rho^3\right],
 \label{eq:ampgl_firstorder_short}
\end{equation}
corresponds to the  asymmetric version of the  Cahn-Hilliard (CH) equation, see e.g. Refs. \cite{Bray:1994.1,Desai:2009},
The coefficients $\alpha_i$ are given in eqs.~(\ref{eq:alpha_g}).
Note that the quadratic nonlinearity implies a broken $\pm\rho$-symmetry.
This is usually not included in the classic representation of the CH equation 
since it can be removed by adding a constant to the density: $\rho\rightarrow \rho+\rho_h$.
In any case, the quadratic nonlinearity vanishes for  $\alpha_3=0$. 
For MIPS, this is fulfilled for $r_1=0$, or $\bar\rho=\rho_*$ accordingly.
This special case has also been considered in \cite{SpeckT:2015.1}
where they found a CH equation with coefficients consistent with $\alpha_i$ above.\\

Equation (\ref{eq:ampgl_firstorder_short}) can be derived from the energy functional
\begin{align}
F=\int \left[-\frac{\alpha_1}{2}+\frac{\alpha_2}{2}(\partial_x\rho)^2- \frac{\alpha_3}{3}\rho^3+\frac{\alpha_4}{4}\rho^4 \right]dx
\end{align}
via
\begin{align}
\partial_t\rho=\partial_x^2\frac{\delta F}{\delta \rho}\,.
\end{align}
On first glance this is a surprising result since
the two initial dynamical equations for the density, eq.~(\ref{eq:density}), and the polarization, eq.~(\ref{eq:polarization}), 
do not follow potential dynamics and therefore cannot be derived from a functional. 
Nevertheless, this specific property has been seen for other non-equilibrium systems:
The evolution equation for the envelope of spatially periodic patterns  
also follows potential dynamics while the dissipative starting equations do not \cite{CrossHo,Cross:2009}.

\subsection{Extended CH model}
\label{sec:discussion_higherorder}

We now take a closer look at the CH model extended to the next higher order,  eq.~(\ref{eq:ampgl_full_short}) with coefficients $\beta_i$ given in eqs.~(\ref{eq:beta_i}).
The contributions $\beta_1$, $\beta_2$ and $\beta_3$ are 
 corrections to the coefficients $\alpha_1$, $\alpha_2$ and $\alpha_3$ of the leading order CH equation.
Note, however,  that according to eqs.~(\ref{eq:beta1}) and (\ref{eq:beta3}), $\beta_1$ and $\beta_3$ are 
functions of $\varepsilon$ and thus both increase with the distance $\varepsilon$ from phase separation onset.
Notably, $\beta_3$ - the correction to the quadratic nonlinearity - is 
not a function of the relative deviation $r_1$ from the critical density parameter $R_*$.
Thus, while for $r_1=0$ the CH model at leading order is $\pm\rho$-symmetric, 
the symmetry is always broken at higher order.

The coefficients $\beta_5$ and $\beta_6$ are the prefactors of higher order nonlinearities.
These new contributions $\propto \partial_x^2 (\partial_x \rho)^2$  and $\propto \partial_x^4 \rho^2$
are structurally different compared to the terms in the leading order CH model.
In general, an additional nonlinearity $\propto\partial_x^2 \rho^4$ is of the same order as these two contributions.
However, in the exemplary case of MIPS we analyze here this term does not appear.
Note, however, that the higher order extension  of the CH model presented 
here can also be applied to other active phase separation systems. 
We expect the additional nonlinearity of the form $\propto\partial_x^2 \rho^4$ to be relevant in other examples 
such as cell polarization or chemotaxis.

In the context of MIPS, a contribution $\propto \partial_x^2 (\partial_x \rho)^2$ 
has been introduced via a phenomenological approach in Ref.~\cite{Cates:2014.1}. 
The CH model extended by this term has been called {\it Active Model B}.
It was considered as a non-equilibrium extension of the CH model and minimal model for MIPS.
We would like to reiterate that the  CH model as given by eq.~(\ref{eq:ampgl_firstorder_short})
(without any additional nonlinear terms) 
is the leading order description of the {\it non-equilibrium} 
phenomenon of active phase seperation \cite{Bergmann:2018.2}. 
As we have shown here, this also includes MIPS.
All higher order nonlinearities vanish for $\varepsilon\rightarrow 0$ 
(see also the discussion in sec.~\ref{sec:rescaling}). 
In that respect {\it Active Model B} is a nonlinear extension of the CH model - not an extension of the CH model to non-equilibrium systems.
Our systematic approach reveals the existence of the additional higher nonlinearity  
$\propto \partial_x^4 \rho^2=2\partial_x^2 [ (\partial_x \rho)^2+\rho\partial_x^2 \rho]$. 
It includes the nonlinear correction to the CH model, $ \propto \partial_x^2  (\partial_x \rho)^2$, that leads to
the {\it Active Model B} \cite{Cates:2014.1,Cates:2015.1}.
The second part of the new nonlinear correction term, $\propto \partial_x^2 ( \rho \partial_x^2 \rho)$,
has recently been included in a further CH extension for MIPS called {\it Active Model B+} \cite{Solon:2018.1,Cates:2018.1}.
Note that the contribution $\propto \beta_6$ in eq.~(\ref{eq:ampgl_full_short}) vanishes for $\lambda=0$.
{\it Active Model B} and {\it Active Model B+} also do not include the  quadratic nonlinearity $\propto \beta_3 \rho^2$. 
Our analysis  shows, however, that the coefficients $\beta_i$ in general
are not independent from each other and $\beta_2$
in fact always  appears simultaneously 
with the nonlinearity $\propto \beta_5$.}
The broken $\pm$-symmetry and the resulting asymmetric phase separation profiles depend 
on the distance $\varepsilon$ from threshold (see $\beta_3$ in eq.~(\ref{eq:beta3}).
It is an important qualitative feature of the system behavior above threshold. 

As discussed in sec.~\ref{sec:discussion_leading}, the leading order CH model can be derived from an energy potential. 
For the extended Ch model, eq.~(\ref{eq:ampgl_full_short}), integrability depends on the coefficients
of the additional higher order contributions:
For arbitrary values of $\beta_5$ and $\beta_6$, the extended CH model is non-integrable. 
In the special case $\beta_6=-\beta_5$, however, eq.~(\ref{eq:ampgl_secondorder}) 
can be derived from the energy functional 
\begin{align}
F=\int &\left[ \frac{-\alpha_1+\beta_1}{2}\rho^2 + \frac{\alpha_2+\beta_2}{2}(\partial_x \rho)^2\right. \nonumber\\
 &\left.-\frac{\alpha_3+\beta_3}{3}\rho^3 -\frac{\alpha_4}{4}\rho^4 +\frac{\beta_5}{2} \rho^2\partial_x^2\rho\right]  dx \,.
\end{align}
For the MIPS model, eqs.~(\ref{sepbasic}), this condition is fulfilled for 
\begin{align}
\lambda^2 = \frac{\zeta {v_\star}^2}{8}\,.
\label{eq:lambda_integrability}
\end{align}
Note, however, that the linear stability analysis in sec.~\ref{sec:linsta} introduced 
a condition for $\lambda$: $\lambda^2<\zeta v_*^2/24$ in eq.~(\ref{eq:lambda_condition}). 
This condition and eq.~(\ref{eq:lambda_integrability}) cannot be fulfilled simultaneously.
Thus, the integrability of the extended CH model depends on the exact parameter choices.
For the MIPS continuum model we investigate here, 
there do not seem to be suitable parameter choices that enable integrability.
But note again that our approach can be applied to other systems
showing active phase separation. 
For these other models,  the coefficients of the extended CH model
could allow for integrability.

\subsection{Comparison of linear stability}
\label{sec:thresholds}

As a first step to assess the quality of our derived reduced equations, 
we analyze the linear stability of the homogeneous basic state $\rho=0$,
and compare to the stability of the full MIPS model.
As discussed in sec.~\ref{sec:linsta}, the instability condition for the full MIPS system 
is given by eq.~(\ref{eq:instability_condition}).
Using $v_0=v_*(1+\varepsilon)$, $R=R*(1+r_1)$ and the definitions of $D_e$ and $R_*$ 
as given by eqs.~(\ref{eq:vstar_def}),
we find
\begin{align}
 \varepsilon_c &=\frac{1}{8}\left(1+9r_1\right)-\frac{1}{8}\sqrt{1+18r_1+9r_1^2}\nonumber\\
 &\approx \frac{9}{2}r_1^2-\frac{81}{2}r_1^3 + \frac{891}{2}r_1^4 + \mathcal{O}(r_1^5)
 \label{eq:threshold_full}
\end{align}
 for the onset of phase separation.
Thus, in the symmetric case $r_1=0$ the threshold is $\varepsilon_c=0$.
For $r_1\neq 0$ the onset of phase separation is shifted to larger values of $\varepsilon$. 
Larger particle velocities $v_0$ are thus required to trigger the demixing process.

Similarly, we can analyze the linear stability of both the leading order CH equation, eq.~(\ref{eq:ampgl_firstorder_short}),
and its higher order extension, eq.~(\ref{eq:ampgl_full_short}).
The threshold calculated from the linear parts of eq.~(\ref{eq:ampgl_firstorder_short}) is given by 
\begin{equation}
\varepsilon_{c,\text{lead}}=\frac{9}{2}r_1^2\,.
\end{equation} 
Comparing this to $\varepsilon_c$ in eq.~(\ref{eq:threshold_full}), we find that the 
shifting of the threshold due to finite $r_1$ is represented up to leading  order of $r_1$.
Assuming $r_1>0$, $\varepsilon_{c,\text{lead}}$ significantly overestimates the real threshold $\varepsilon_c$.
For the extended CH equation, eq.~(\ref{eq:ampgl_full_short}), we find the threshold
\begin{align}
 \varepsilon_{c,\text{ext}} &=\frac{9r_1^2}{2\left(1+9r_1^2\right)}\nonumber\\
 &\approx \frac{9}{2}r_1^2-\frac{81}{2}r_1^3 + \frac{729}{2}r_1^4 + \mathcal{O}(r_1^5).
\end{align}
This is in agreement with the threshold for the full model, eq.~(\ref{eq:threshold_full}), up to the order $\mathcal{O}(r_1^3)$.
The threshold is therefore only slightly underestimated compared to the full model.
Keeping these different threshold values in mind is particularly important for the numerical comparison of the 
MIPS model, eqs.~(\ref{eq:density}) and (\ref{eq:polarization}), 
to its two reductions, eqs.~(\ref{eq:ampgl_firstorder_short}) and (\ref{eq:ampgl_full_short}) in sec.~\ref{sec:numerics}.
All three equations only provide the exact same threshold, namely $\varepsilon_c=0$, in the special case $r_1=0$.\\

The linear stability analysis also provides the dispersion relation for the perturbation growth rate $\sigma$.
For the full model, it is given by eq.~(\ref{basicdisp}).
Expanding for small perturbation wavenumbers $q$, the general form of the growth rate is
\begin{equation}
\sigma = D_2q^2 - D_4q^4 + {\cal O}(q^6)\,.
\end{equation}
The coefficients $D_2$ and $D_4$ are given in eqs.~(\ref{eq:full_D2}) and (\ref{eq:full_D4}), respectively.
Using the definitions introduced in the course of the perturbative expansion, $D_2$ can be rewritten to 
\begin{equation}
 D_2 = \frac{1}{8}v_*^2\varepsilon - \frac{9}{16}v_*^2r_1^2 + \frac{9}{8}v_*^2r_1\varepsilon - \frac{1}{2}v_*^2\varepsilon^2\,.
\end{equation}
Good agreement between the full MIPS model and its reduction to eq.~(\ref{eq:ampgl_full_short})
can only be expected if the reduced equations are able to reproduce the basic form of this growth rate. 
The linear part of eq.~(\ref{eq:ampgl_full_short}) leads to a growth rate of the form 
\begin{equation}
 \sigma(q)=G_2q^2-G_4q^4,
\end{equation}
where
\begin{align}
 G_2 &= \frac{1}{8}v_{*}^2\varepsilon-\frac{9}{16}v_{*}^2r_1^2+\frac{9}{8}v_{*}^2 r_1\varepsilon,\\
 G_4 &= \frac{1}{256}v_{*}^4-\frac{3}{32\zeta}\lambda^2v_{*}^2+\frac{3}{16\zeta}\lambda^2v_{*}^2r_1.
\end{align}
$G_2$ is in agreement with $D_2$ of the full model equations up to linear order in $\varepsilon$. 
$D_2$ only includes an additional term of order $\mathcal{O}(\varepsilon^2)$: $D_2 = G_2 - v_*^2\varepsilon^2/2$.
$G_4$ exactly reduces to $D_4$ in the case $\varepsilon=r_1=0$. 
In the limit $\varepsilon\rightarrow 0$ but $r_1\neq 0$, the two terms agree up to linear order in $r_1$.
As discussed in sec.~\ref{sec:linsta}, the coefficient $D_4$ has to be positive 
for the instability condition to hold and to ensure damping of short wavelength perturbations.
The same applies to the coefficient $G_4$. 
The condition $G_4>0$ is fulfilled if 
\begin{equation}
 \lambda^2<\frac{1}{24}v_*^2\zeta\frac{1}{1-2r_1}.
 \label{eq:lambda_condition2}
\end{equation}
Note the similarity to the previously derived condition in eq.~(\ref{eq:lambda_condition}).

\subsection{Significance of nonlinear corrections}
\label{sec:rescaling}

In this section, we discuss the importance of the higher order nonlinearities
compared to the leading order terms  of the classic Cahn-Hilliard model in eq.~(\ref{eq:ampgl_firstorder_short}). 
For this comparison we focus on the case with $\pm$-symmetry at leading order, {\it i.e.} $r_1=0$.
We rescale time, space and amplitude in eq.~(\ref{eq:ampgl_full_short})
via $t'=\tau_0 \varepsilon^2 t$, $x'=\xi_0 \sqrt{\varepsilon} x$ and 
$\rho'=\rho_0\rho/\sqrt{\varepsilon} $, respectively, where
\begin{subequations}
\begin{align}
 \tau_0 &= \frac{4\zeta v_*^2}{v_*^2\zeta-24\lambda^2}\,,\\
 \xi_0^2 &= \frac{32\zeta}{v_*^2\zeta-24\lambda^2}\,,\\
 \rho_0 &= \frac{2\sqrt{6}}{3}\frac{\zeta}{v_*}\,.
\end{align}
\end{subequations}
This allows us to rewrite eq.~(\ref{eq:ampgl_full_short}) in the following form:
\begin{align}
 \partial_{t'}\rho' =& -\partial_{x'}^2\left[ \rho'+\partial_{x'}^2\rho'-\rho'^3 \right]\nonumber\\
 &-  \sqrt{\varepsilon} \partial_{x'}^2  \left[\gamma_1\rho'^2 +\gamma_2\partial_{x'}^2\rho'^2  + \gamma_3 \left(\partial_{x'}\rho'\right)^2\right]\,,
\label{eq:amp_rescaled}
 \end{align}
where 
\begin{subequations}
\begin{align}
 \gamma_1 &= \frac{3\sqrt{6}}{2}\,,\\
 \gamma_2 &= \frac{8\sqrt{6}\lambda^2}{v_*^2\zeta-24\lambda^2}\,,\\
 \gamma_3 &= \frac{\sqrt{6}\left(3v_*^2\zeta-40\lambda^2\right)}{2\left(v_*^2\zeta-24\lambda^2\right)}\,.
\end{align}
\end{subequations}
The first line in eq.~(\ref{eq:amp_rescaled}) is the parameter-free, $\pm\rho$-symmetric version of the Cahn-Hilliard model as 
described, {\it e.g.}, in Refs.~\cite{Bray:1994.1,Desai:2009}.
The  additional three contributions are the first higher order corrections as
gained above via a systematic reduction of the continuum model for MIPS.
These three corrections are proportional to $\sqrt{\varepsilon}$ and thus vanish 
when approaching  the onset of active phase separation ($\varepsilon \to 0$).
In the limit $\varepsilon \to 0$ the classic CH model thus fully describes the non-equilibrium mean-field  dynamics of  MIPS.
 With increasing $\varepsilon$, the higher order contributions become more and more important.

Note that eq.~(\ref{eq:amp_rescaled}) was derived under the assumption $r_1=0$.
As discussed in sec.~\ref{sec:discussion_leading}, the CH model at leading order is $\pm\rho$-symmetric in this case.
The  three higher order contributions in eq.~(\ref{eq:amp_rescaled}), however, break the $\pm\rho$-symmetry with increasing $\varepsilon$.
Moreover, in the case of the MIPS model we analyze here, 
the coefficient $\gamma_1$ does not depend on any of the system parameters at all.
Thus, there is in fact no special case in which this contribution can be neglected.

The coefficients of the other two higher order nonlinearities, $\gamma_2$ and $\gamma_3$, 
are functions of the system parameters, especially of $\lambda$.
Typical parameter choices for the continuum  model in eq.~(\ref{sepbasic})
are such that  $v_*$ and $\zeta$ are of order $\mathcal{O}(1)$. 
Accordingly, $\lambda$ has to be small to fulfill the condition in eq.~(\ref{eq:lambda_condition}).
Therefore, an expansion of $\gamma_2$ and $\gamma_3$ 
 in terms of small $\lambda$ is appropriate:
\begin{align}
 \gamma_2 &= \frac{8\sqrt{6}}{v_*^2\zeta}\lambda^2 + \mathcal{O}(\lambda^4)\,,\label{eq:gam2}\\
 \gamma_3 &= \gamma_1+2\gamma_2 +   \mathcal{O}(\lambda^4) \,. \label{eq:gam3} 
\end{align}
In the limit $\lambda=0$ the coefficient $\gamma_2$ vanishes, {\it i.e.} $\gamma_2=0$,
and $\gamma_3$ simplifies to $\gamma_3=\gamma_1$.
For finite $\lambda$, $\gamma_2$ also becomes finite.
But since according to eq.~(\ref{eq:gam2}) $\gamma_2$ is proportional to $\lambda^2$, 
it  will be much smaller than $\gamma_3$ for small $\lambda$.
For MIPS as described by the mean-field model in eqs.~(\ref{sepbasic}),
the impact of the nonlinearity $\propto\partial_x^2(\partial_x\rho)^2$ thus seems 
to overshadow the term $\propto\partial_x^4\rho^2$.
This predominance of $\gamma_3$, however, is specific to MIPS. 
For other examples of active phase separation such as cell polarization or chemotactically communicating cells, 
 we expect that the nonlinearities described by $\gamma_1$ or $\gamma_2$ can be of similar order as $\gamma_3$.
As mentioned earlier, for both examples of active phase separation we also expect an additional  
 higher order correction $\propto\partial_x^2\rho^4$ which is completely absent in the MIPS model.

\section{Numerical comparison}
\label{sec:numerics}

In this section, we compare numerical simulations of the full MIPS model, eqs.~(\ref{sepbasic}),
to both the leading order CH equation, eq.~(\ref{eq:ampgl_firstorder_short}), 
as well as the extended version including higher nonlinearities, eq.~(\ref{eq:ampgl_full_short}).
On the one hand, this allows us to assess the quality and validity range of our reduction scheme in general.
On the other hand, comparing the leader order and the extended CH model 
also gives us information about the importance of higher order nonlinearities in MIPS.

All simulations were performed using a spectral method with a semi-implicit Euler time step.
The system size was $L=100$ with periodic boundary conditions and $N=256$ Fourier modes were used.

\begin{figure}
 \includegraphics[width=\columnwidth]{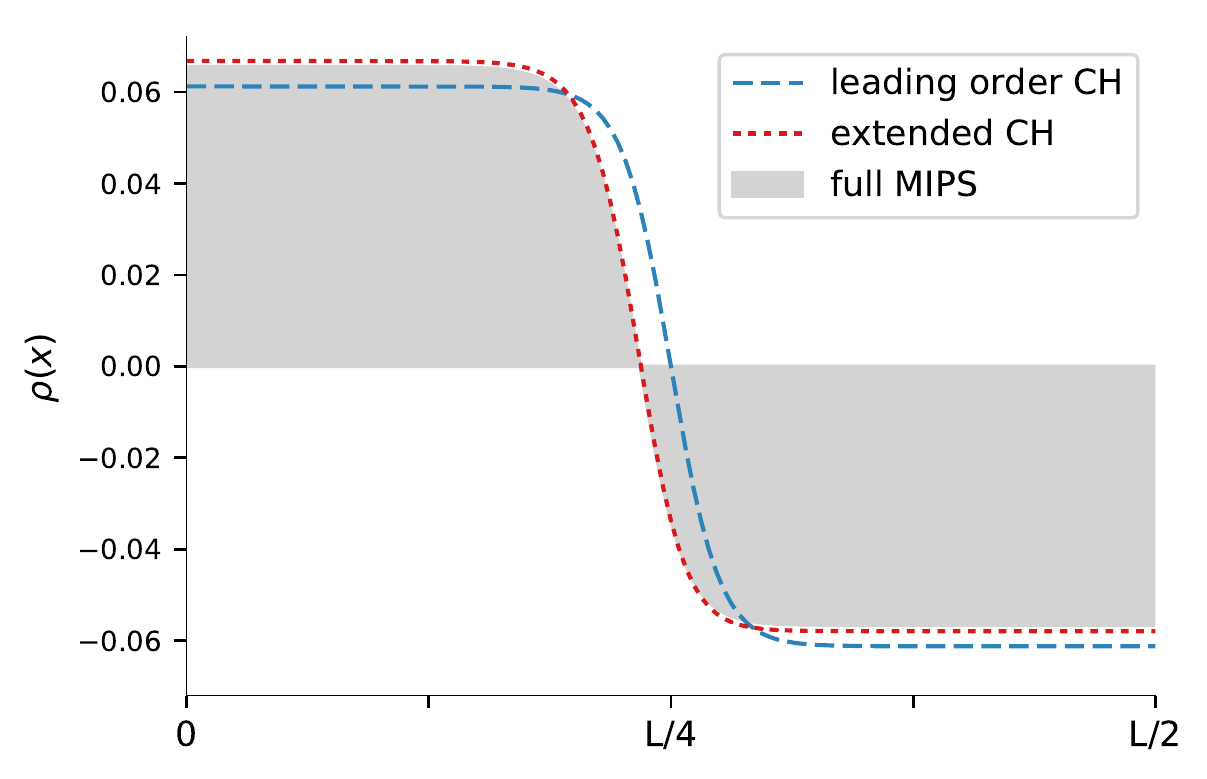}
 \caption{Comparison of the steady state profiles in the ``symmetric'' case ($\bar\rho=\rho_*$)
 at $\varepsilon=0.01$: 
 full MIPS model (shaded grey) vs leading order CH equation (blue dashed) 
 vs extended CH equation (red dotted).
 Other parameters: $\zeta=v_*=1$.}
 \label{fig:profiles_symmetric}
\end{figure}

We first analyze the special case $r_1=0$, {\it i.e.} $\bar\rho=\rho_*$. 
This is the case in which the $\pm$-symmetry-breaking quadratic nonlinearity vanishes at leading order.
We choose $v_*=1$ and $\zeta=1$ throughout all of the following simulation results.
As discussed in sec.~\ref{sec:rescaling}, $\lambda$ has to be small and is thus not 
expected to significantly influence the results. We thus set $\lambda=0$.

Figure~\ref{fig:profiles_symmetric} shows the steady state profiles for the three models
(full MIPS model, leading order CH and extended CH) at $\varepsilon=0.01$. 
The profiles are typical for phase separation solutions:
We find two distinct regions where the mean density is either increased ($\rho>0$) or decreased ($\rho<0$).
In each of the regions $\rho$ is essentially spatially constant,
creating two distinct density plateaus $\rho_{\text{min}}$ and $\rho_{\text{max}}$.
The two plateaus are smoothly connected at their boundary, resembling a hyperbolic tangent function.
Note that the mean density in the system is conserved. 
Thus, the areas under the positive and negative parts of $\rho(x)$ are equal.

The solution for the full system is represented as the outline of the grey shaded area.
We first compare this to the leading order CH equation (blue dashed line).
As predicted, the leading order CH equation results in a symmetric phase separation profile,
{\it i.e.} the two plateaus have the same absolute value: $\rho_{\text{max}}=\lvert\rho_{\text{min}}\rvert$.
This does not accurately represent the solution for the full system, which is already slightly asymmetric.
However, the leading order CH equation gives a good approximation of the plateau values 
with a deviation of less than $7\%$ from the real value.
Extending the CH equation to the next higher order (red dotted line in fig.~\ref{fig:profiles_symmetric}),
we can almost perfectly reproduce the profile for the full MIPS model. 
It accurately represents the asymmetry of the phase separation profile.
The deviation in the plateau values shrinks to less than $2\%$.\\

\begin{figure}
 \includegraphics[width=\columnwidth]{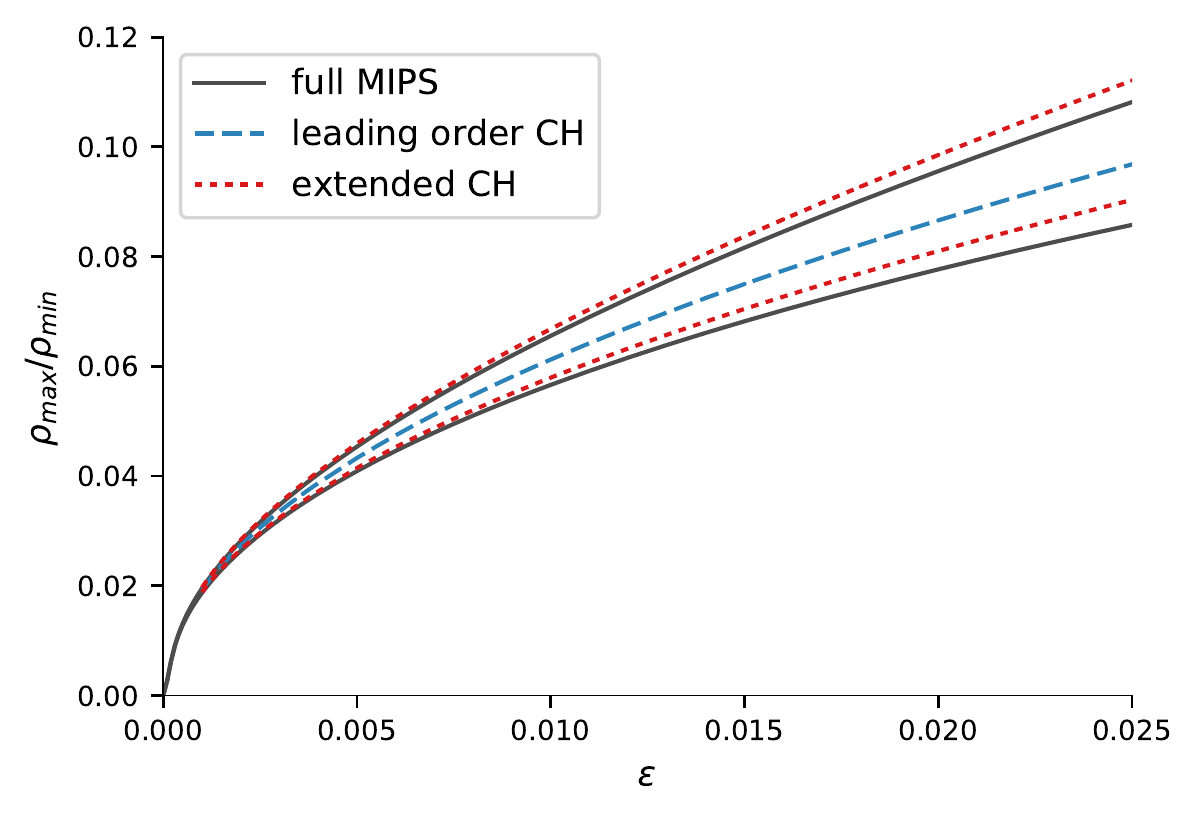}
 \caption{Comparison of plateau values $\lvert\rho_{\text{min}}\rvert$ and $\rho_{\text{max}}$ 
 as a function of the control parameter $\varepsilon$ for $\bar\rho=\rho_*$ ({\it i.e.} $r_1=0$):
  full MIPS model (black solid) vs leading order CH equation (blue dashed) 
 vs extended CH equation (red dotted).}
 \label{fig:ampeps_symmetric}
\end{figure}

Figure~\ref{fig:ampeps_symmetric} shows the absolute plateau values $\lvert\rho_{\text{min}}\rvert$ and $\rho_{\text{max}}$ 
as a function of $\varepsilon$ - the distance from phase separation onset.
The bifurcation to active phase separation is supercritical in this case:
starting at $\varepsilon_c=0$, the plateau values increase monotonically.
Considering only the leading order approximation (blue dashed line), we again find 
the system to be symmetric for all values of $\varepsilon$.
In reality, the full system (black solid lines) becomes more and more asymmetric for increasing $\varepsilon$.
This is very accurately represented by the higher order approximation (red dotted lines).
It only starts to deviate from the full model further from threshold.
Importantly though, close to the onset of mobility induced phase separation, as $\varepsilon$ becomes smaller, 
the full model becomes more and more symmetric.
All three models then are in increasingly good agreement. 
This again underlines the fact that the classic CH model is the simplest generic model for active phase separation.
All active phase separation phenomena of this type can be reduced to the CH model close to onset.
Higher order nonlinearities only come into play further from threshold.

\begin{figure}
 \includegraphics[width=\columnwidth]{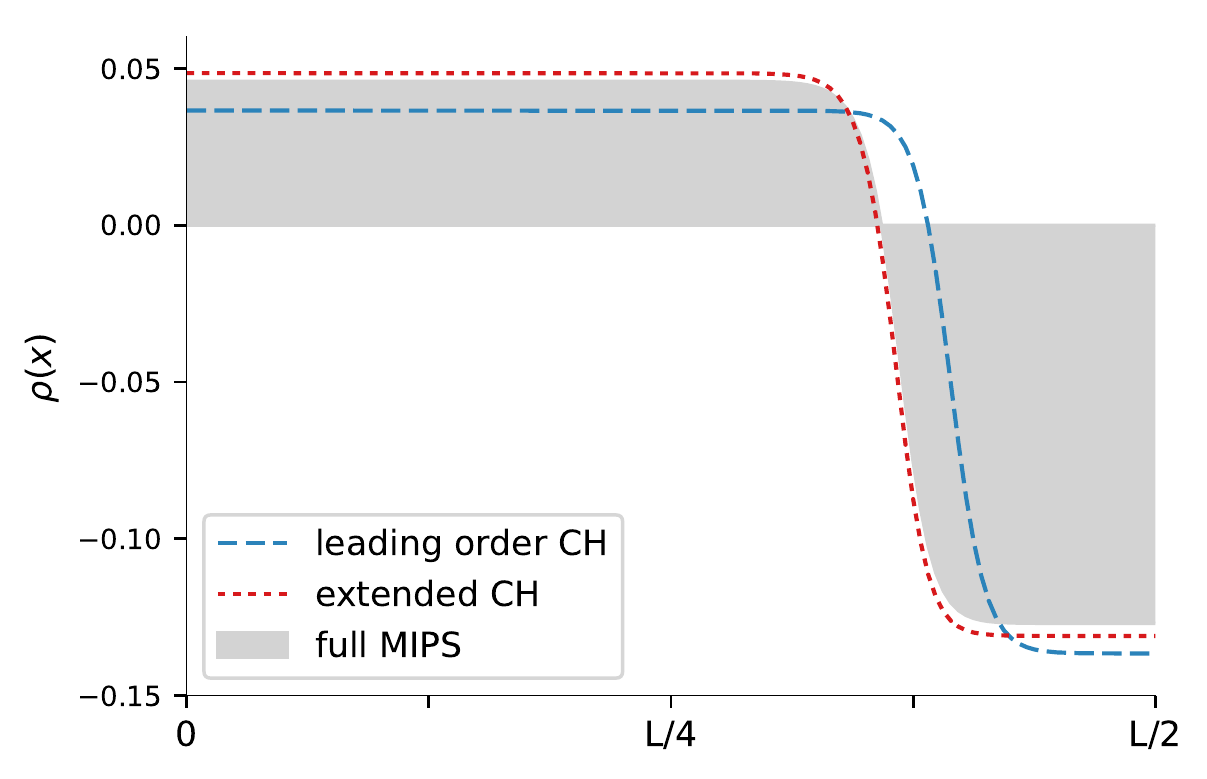}
  \caption{Comparison of the steady state profiles for $\bar\rho=0.8$ at $\varepsilon=0.02$: 
  full MIPS model (shaded grey) vs leading order CH equation (blue dashed) 
 vs extended CH equation (red dotted).}
 \label{fig:profiles_asymmetric}
\end{figure}

If we allow $r_1\neq0$, phase separation is asymmetric even at leading order.
This can be seen in fig.~\ref{fig:profiles_asymmetric} which shows the steady state profiles for the full MIPS model,
leading order CH and extended CH at $\varepsilon=0.02$. 
Here, the leading order CH equation (dashed blue line) results in an asymmetric solution.
However, the predicted plateau values deviate about $20\%$ from the full system (outlines of shaded grey region). 
The extended CH model, meanwhile, is still able to accurately predict the full system solution
with a deviation of less than $6\%$. 

\begin{figure}
 \includegraphics[width=\columnwidth]{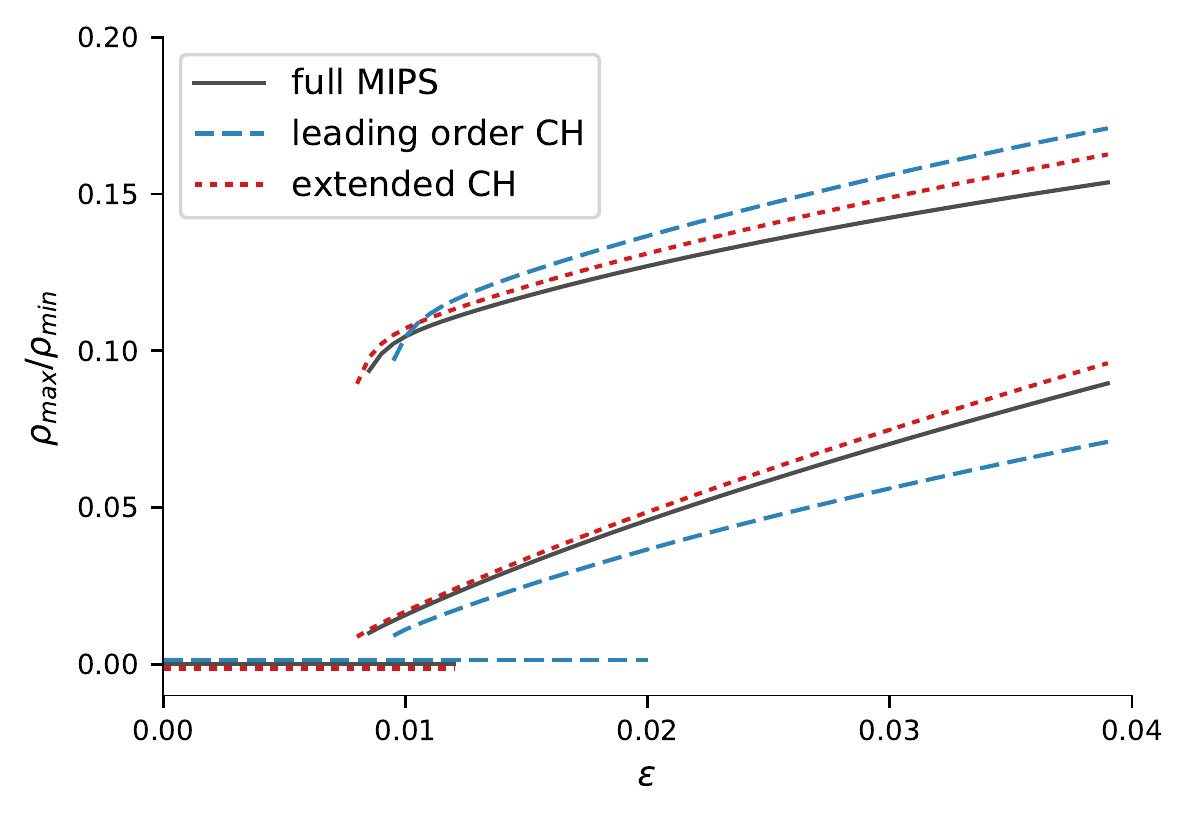}
 \caption{Comparison of plateau values $\lvert\rho_{\text{min}}\rvert$ and $\rho_{\text{max}}$ 
 as a function of the control parameter $\varepsilon$ for $\bar\rho=0.8$ (or $r_1=1/15$):
  full MIPS model (black solid) vs leading order CH equation (blue dashed) 
 vs extended CH equation (red dotted).}
 \label{fig:ampeps_asymmetric}
\end{figure}

Looking at the plateau values as a function of $\varepsilon$ (see fig.~\ref{fig:ampeps_asymmetric})
solidifies this impression:
The leading order CH model gives a good qualitative representation of the full system.
Going to the extended CH model provides very good quantitative agreement with the full model 
even for larger values of $\varepsilon$.
As discussed earlier in sec.~\ref{sec:thresholds}, the onset of phase separation
({\it i.e.} the $\varepsilon$-value at which the homogeneous solution $\lvert\rho_{\text{min}}\rvert=\rho_{\text{max}}=0$
becomes unstable)
is shifted to finite values of $\varepsilon$ in the case $r_1\neq0$.
For the given system parameters, the threshold for the full system is shifted to $\varepsilon_c\approx0.013$.
The leading order CH model significantly overestimates this threshold, shifting to $\varepsilon_c\approx0.02$.
The extended CH model only very slightly underestimates the real threshold.
Note that above this threshold, the plateau values immediately jump to finite values.
Thus, the transition from the homogeneous to the phase-separated state is no longer smooth.
On the other hand, fig.~\ref{fig:ampeps_asymmetric}) also shows that the branches
of finite density plateau values extend below the thresholds noted above.
This creates a range of bistability - 
a range of control parameter values in which both the homogeneous and the phase-separated state
are stable simultaneously.
All of these characteristics indicate that bifurcation from 
the homogeneous state to active phase separation is now subcritical.

%%%%%%%%%%%%%%%%%%%%%%%%%%%%%%%%%%%%%%%%%%%%%%%%%%%%%%%%%%%%%%%%%%%%%%%%%%%%%%%
%      SECTION:  conclusion
%%%%%%%%%%%%%%%%%%%%%%%%%%%%%%%%%%%%%%%%%%%%%%%%%%%%%%%%%%%%%%%%%%%%%%%%%%%%%%%

\section{Conclusion}

Starting from the mean-field model for active Brownian particles  in Refs.~\cite{SpeckT:2013.1,SpeckT:2015.1},
we applied a perturbative approach
introduced in Ref.~\cite{Bergmann:2018.2}. We showed 
that the non-equilibrium phenomenon motility-induced phase separation (MIPS)  
is described near its onset at leading order by the Cahn-Hilliard (CH) model \cite{Cahn:58.1,Cahn:1961.1,Bray:1994.1,Desai:2009}.
This is in agreement with a recent observation 
that the CH model  describes the system-spanning behavior of a number of very different demixing phenomena in active and living systems far from thermal equilibrium \cite{Bergmann:2018.2}.
The results in this work show that MIPS also belongs to this class of active phase separation. 
Thus, even though the CH model was originally introduced to describe phase separation of 
binary mixtures in thermal equilibrium,
our analysis shows that it is also the generic leading order description of active phase separation 
- a non-equilibrium phenomenon.
% Second, the CH equation can be derived from a functional.
% The initial mean-field equations for active Brownian particles, however, 
% do not follow potential dynamics. 
% This specific feature of active phase separation is shared by 
% numerous other non-equilibrium phase transitions (bifurcations).
% The evolution equation for the envelope of spatially periodic patterns, {\it e.g.}, 
% also follows potential dynamics while the dissipative starting equations do not \cite{CrossHo,Cross:2009}.

We also extended the perturbative scheme introduced in Ref.~\cite{Bergmann:2018.2} beyond the CH model
to next higher order nonlinearities.
In this work, we used the continuum model for MIPS as a framework to establish this concept. 
The extension of our nonlinear expansion, however, can also be applied
 to other systems showing active phase separation (with a conserved order parameter field) 
 such as cell polarization and clustering of chemotactically communicating cells.
Having a $\pm$-symmetric CH model at onset of active phase separation,
we find  that in general four nonlinear terms come 
into play at the next higher order.
Two of them have the same form as contributions suggested
in previous phenomenological extensions of the CH model for MIPS \cite{Cates:2014.1,Cates:2015.1,Solon:2018.1,Cates:2018.1}.
 These phenomenological models are thus related to the extended CH model that
 our perturbative scheme provides.
 Our approach, however, is non-phenomenological: It establishes a direct mathematical link
 between the coefficents of the extended CH model and the full mean-field description of MIPS
 (or any other basic model of active phase separation in general). 
 It shows in addition, that the  coefficients of the additional contributions  in the extended CH model 
 are in general not independent from each other, as often assumed in phenomenological
 approaches. Furthermore, these coefficents are system-specific 
 and cannot be removed by rescaling as in the case of the leading order CH model.
 It is also important to reiterate that these nonlinear extensions become negligible 
 when approaching the onset of MIPS or other examples of active phase separation.
 Therefore, the leading order CH model already covers the universal behavior of MIPS (as a non-equilibrium phenomenon) 
 near its onset. 
 Higher order nonlinearities mainly improve accuracy and become relevant further from threshold.
 They should thus not be seen as the key to expand the CH model to non-equilibrium systems.

Within the systematics of pattern formation theory, 
the work we introduced in Ref.~\cite{Bergmann:2018.2} and extended here 
is a weakly nonlinear analysis and reduction method for active phase separation described 
by conserved order parameter fields.
It can be seen as a yet unexplored counterpart to
the  weakly nonlinear analysis of (non-oscillatory) spatially periodic patterns with unconserved order parameter fields
and its numerous applications \cite{CrossHo,Pismen:2006,Cross:2009,Meron:2015,Misbah:2016}.

Our generic approach for active phase separation opens up several pathways for further system-spanning investigations.
Coarsening dynamics in large systems, and especially the role of higher nonlinearities in this context, 
have already been of particular interest to the scientific community (see, {\it e.g.}, Ref.~\cite{Cates:2018.1}  for MIPS).
Other active phase separation phenomena such as cell polarization, on the other hand, take place in very small 
systems where coarsening plays a less important role \cite{EdelsteinKeshet:2013.1}. 
For these systems, spatial constraints may significantly influence the behavior instead.
Studies on spatially periodic patterns have already shown that confinement  may trigger 
various interesting generic effects (see {\it e.g.} \cite{Rapp:2016.1}) 
and even induce patterns in small systems which are unstable in larger systems 
(\cite{Bergmann:2018.1} and references therein).
On the basis of our results, it will be interesting to investigate finite size effects
on non-equilibrium phase transitions with conservation constraints.

\begin{acknowledgement}
  Support by the Elite Study Program Biological Physics is gratefully acknowledged.
\end{acknowledgement}

\section*{Author Contribution Statement} 
All authors contributed to the design of the research, to calculations, the interpretation of results 
and the writing of the manuscript. LR performed numerical simulations.

\section*{Data vailability statement}
The simulation datasets used in this article are available from the corresponding author on request.

\bibliographystyle{unsrt}
%\bibliographystyle{eplbib}
% \bibliography{../../../../bib/pattern,../../../../bib/bio,../../../../bib/colloid,../../../../bib/stokes}
%\bibliography{PNASLit/pattern,PNASLit/bio,PNASLit/poly2}

\end{document}